**Title:** Using probability and rules of interaction to simulate the spin relaxation in a MRI

**Author:** John Laurence Haller Jr.

**Affiliation:** Stanford University

**Abstract:** A computer code is written that simulates the relaxation back to thermal equilibrium of an ensemble of particle spins after a pi/2 pulse. Beginning with Bloch's equations the exponential relaxation behavior is discussed and the transition into a step by step process (from the continuous process) is made such that it is possible for the computer code to simulate the action. An analysis of Boltzmann's factor is offered and serves as the link between the temperature of the ensemble and the parameter that is used to determine if a spin is in the up state or the down state. The specifics of the code are discussed and the spins are shown to follow the rules of interaction governed by the physical collisions that can take place. Lastly, graphs of the simulation are provided and a discussion of its usefulness is given.

**Introduction:** A model is provided that uses conditional statements to determine if a random event occurs. The random event is simulated by generating a random number between zero and one and comparing it against a Parameter, alpha1, alpha2, or alpha3. If the random number is smaller than the parameter, the random event is simulated and the model implements the consequences of that event.

Three types of events are simulated: one is a spin spin collision, a second is the spin lattice collision and the third is a precession event. In all three types, conditional statements are used to determine if an individual particle's spin should be rotated. The result is a model of Bloch's equation for nuclear precession and relaxation.

$$\frac{dM}{dt} = M \times gB - \frac{M_x + M_y}{T_2} - \frac{M_z - M_o}{T_1}$$

The Bloch equation gives a phenomenological solution to spin relaxation and precession. However the Bloch equation looks at the macroscopic problem as a whole, and not from the individual spin dynamics. Bloch's equation represents the average value while the model provides the information for the individual spins, including a variance. Thus the model is a more thorough representation of the natural relaxations to thermal equilibrium. Further the model provides the bounded area of possible spin orientations that are left unidentified by Bloch's equation.

Below is the Matlab computer code that simulates how nuclear spins act in a constant magnetic field. The difficulty in creating a complete model is fitting the Parameters (alpha1, alpha2, and alpha3) to each other so that a couple of stable oscillations occur in the mean time of the exponential decay, t2. Currently the model provides an accurate (if the number of particles, N, is large enough) representation for transverse decay and longitudinal growth by simulating the probability of a spin lattice and spin spin collisions represented in the Parameters, alpha1 and alpha2. The results of the model resemble the Bloch equation without the cross product term.

$$\frac{dM}{dt} = -\frac{M_t}{T_2} - \frac{M_z - M_o}{T_1}$$

**Specifics:** The computer model is divided into four unique non-embedded "for" loops each inside a larger "for" loop that represents the time step. The four smaller unique loops simulate all of the particles for each of the four cases of spin up and spin down for both transverse and longitudinal directions. Thus once the simulated spins are separated into one of the four alternatives, the possible collisions that can occur are modeled. These alternatives simulate the longitudinal spin up/down and the transverse spin up/down separately to emulate nature, which does not allow some collisions, for example the spin spin collision does not affect the spins in the longitudinal direction. Once the spins are separated the model compares a random number against the Parameters, alpha1 or alpha2. If a collision is simulated, then, depending upon whether the simulated collision was either a spin lattice collision or a spin spin collision, the spin is either rotated into the longitudinal plane, or it is randomly flipped with probability ½.

In the model, after all the possible collisions of all the particles are simulated, the time step increments. The variables that represent the number of particles in each of the four unique cases are updated to represent the change that occurred as a result of the previous collisions in the last time step. Once the new values for the variables are stored the four unique non-embedded "for" loops are repeated.

In the specific event that the model is looping through either the positive or negative spins in the longitudinal plane, the only collision that can occur is the spin lattice collision. This emulates a consequence of physics. If a collision is simulated, it knocks out a spin from the direction it is in, and replaces that spin in the new direction, selected by comparing a random event against the parameter beta, which is biased towards the side of the thermal equilibrium value. The model emulates the thermal equilibrium that is caused by a constant magnetic field and the Boltzmann factor giving a preference to the lower energy state of the spin aligned with the magnetic field.

**Boltzmann Factor:** The distribution of the probability that a spin will align itself with the spin up state after a collision as a function of the energy difference of the spin up and spin down state, ΔE, is the Fermi distribution.

To determine the probability as a function of the energy difference, ΔE, is to simultaneously solve the three equations that must be satisfied. The first of these equations is that the total number of particles, N, is equal to the number of particles that are spin up, N+, plus the number of particles that are spin down, N–. The second equation is that the difference between the number of spins in the positive direction, N+, minus the number of spins in the negative direction, N-, is equal to the total number of particles, N, times 2β-1, where β is the probability that a spin is placed into the positive spin direction after a collision. This second equation is from the Binomial theorem; note that if β=1, then the difference is N and if β=0, the difference is -N which would imply all the spins are spin up or spin down respectively.

$$N_+ + N_- = N$$

$$N_+ - N_- = N(2\beta - 1)$$

The third equation, the Boltzmann factor, states that the ratio of the number of particles that are spin up to the number of particles that are spin down is equal to the natural exponent raised to the power of the energy difference, ΔE, divided by Boltzmann constant times the temperature, $k_B T$.

$$\frac{N_+}{N_-} = \exp(\Delta E / k_B T)$$

Substituting the first equation for the number of particles into the second equation and dividing by the first equation gives us β as a function of the third equation.

$$\frac{N_+ - N_-}{N_+ + N_-} = \frac{\frac{N_+}{N_-} - 1}{\frac{N_+}{N_-} + 1} = 2b - 1$$

$$b = \frac{\frac{N_+}{N_-}}{\frac{N_+}{N_-} + 1} = \frac{\exp(\Delta E / k_B T)}{\exp(\Delta E / k_B T) + 1} = \frac{1}{\exp(-\Delta E / k_B T) + 1}$$

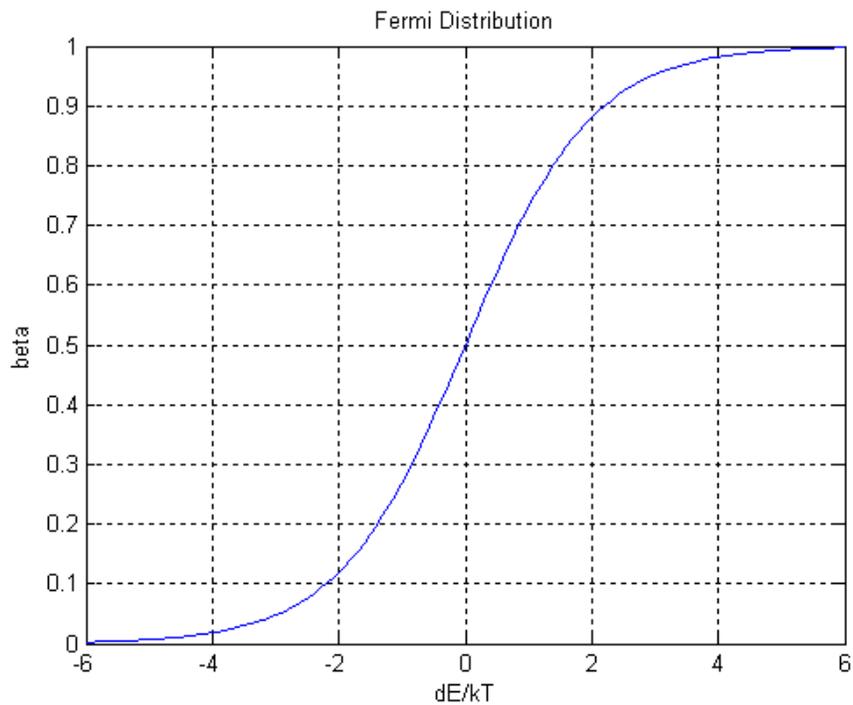

Fermi Distribution

This cumulative distribution is known as the Fermi distribution and has the meaning of the mean occupation number for a Fermion at a given energy. This is consistent with our model because the correct distribution must represent a boolean basis that is either spin up or spin down.

**Collisions:** Using the Fermi distribution as the distribution on the spin state in the longitudinal plane, the model emulates a spin lattice collision. The model deals with spins in the longitudinal and transverse plane that are able to collide with the lattice. If this type of collision is simulated, then the number of particles for that unique pre-collision spin is decreased and the spin is rotated into the longitudinal plane, incrementing the number of spins in the longitudinal plane with the bias given towards spin up as given by the Fermi distribution. In the event that the spin is in the transverse plane, the model is also able to deal with spin spin collisions that also take place. This collision knocks the spin into either the positive or negative direction with equal probability, staying in the transverse plane.

To summarize, the model initializes itself with all the spins in the transverse plane, with β*N spins in the spin up state and (1-β)*N spins in the spin down state. This simulates a pi/2 pulse that would take the asymptotic values of the longitudinal spin states and rotate them into the transverse plane. Once the spins are initialized, the model steps through all the spins and rotates the spins back into the longitudinal plane when a spin lattice collision is simulated. It cycles through all N spins for T time steps. The model conditions on one of four unique outcomes: spin up in the transverse plan, spin down in the transverse plane, spin up in the longitudinal plane, or spin down in the longitudinal plane. Once it is conditioned on the correct unique spin orientation for that particular spin of the total N spins, it samples a random number and compares it against the parameter, alpha1 (for a spin lattice collision) or alpha2 (for a spin spin collision). If the random number is smaller than the parameter a collision event is simulated and the spin is rotated into the direction it should go into given the type of collision that occurred.  Lastly the model randomly determines if the spin is up or down in that orientation.  The model chooses which direction by sampling a second random number and either comparing it against ½ in the event that the spin is in the transverse plane with no magnetic field providing a bias direction, or against beta, β, if the spin is oriented along the longitudinal plane where the magnetic field gives a potential energy savings towards the spin up direction.

**Graphs:**     The output of the model is shown in two graphs.  One is the net magnetization in the transverse plane and the other is the net magnetization in the longitudinal plane.  Those two graphs are below and give the correct exponential rise and fall.  The net transverse number of spins starts at 2000 because this is the net bias that is the asymptotic value that resides in the longitudinal plane.  Since the time sequence starts with a π/2 pulse, the net number of spins, that were in the longitudinal direction, are orientated in the transverse direction and visa versa.  This is why there are zero net spins in the longitudinal direction at t=0.

Note that the value of 2000 comes from a β=.6, and N=10,000.  This is true because (2β-1)N = 2000.  One can note that the exponential rise and fall are slightly different in their rate.  This is fundamentally because T2 is always smaller than T1 because T2 includes both kinds of collisions where T1 relaxes only because of spin lattice collisions.  Also the parameters alpha1 and alpha 2 can be "dialed" into the model as noticed in the next pair of graphs when the parameters were changed.

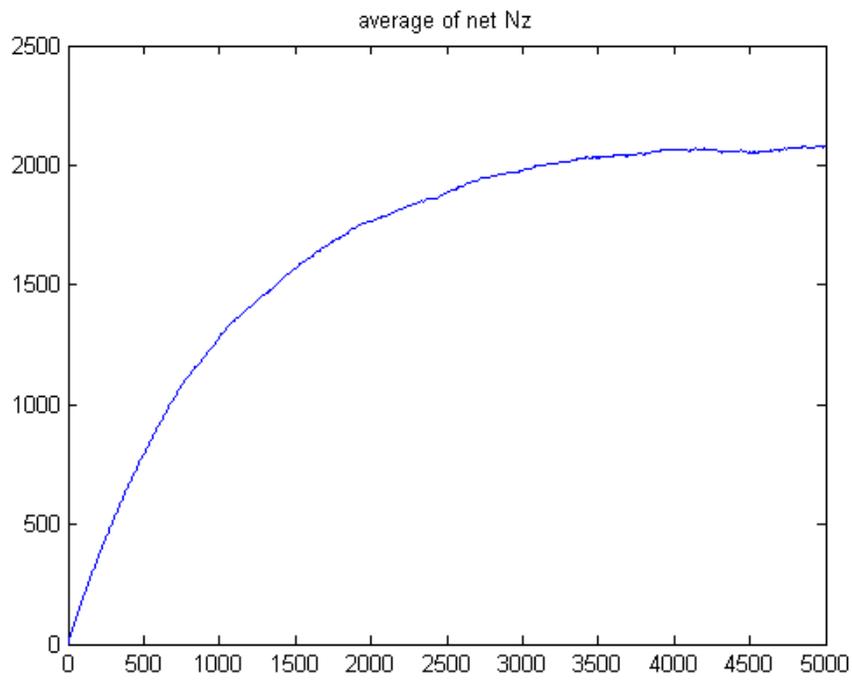

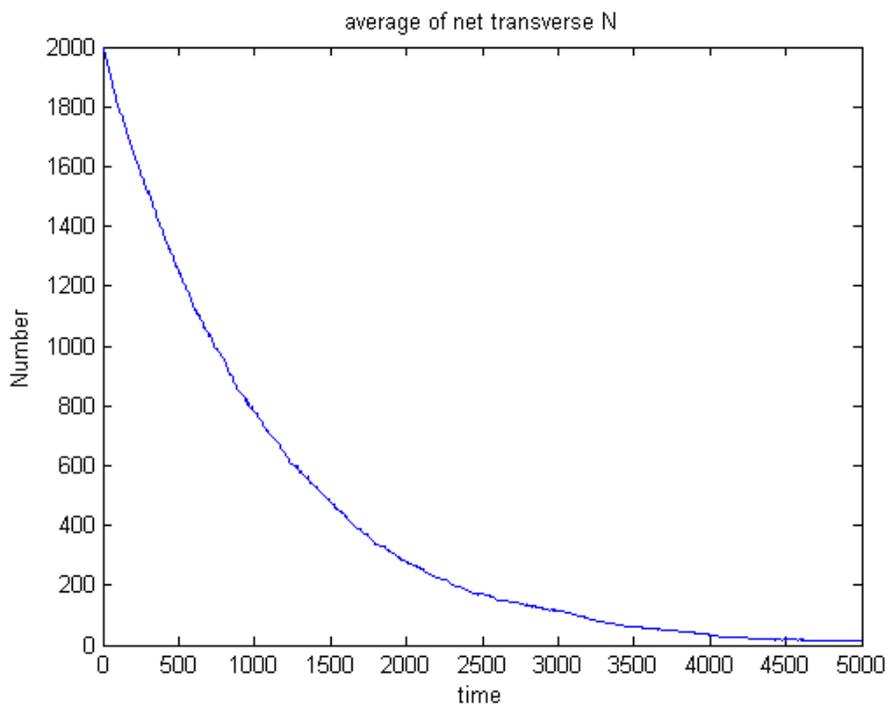

Here are two more graphs with a pi/2 pulse put at t=5000. Notice that the decay parameter for the transverse plane is shorter.

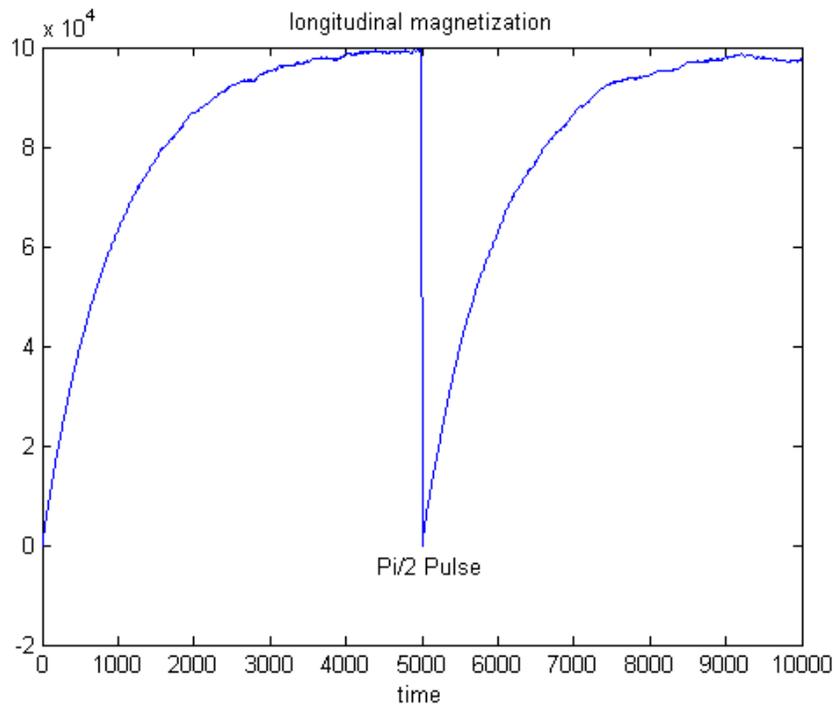

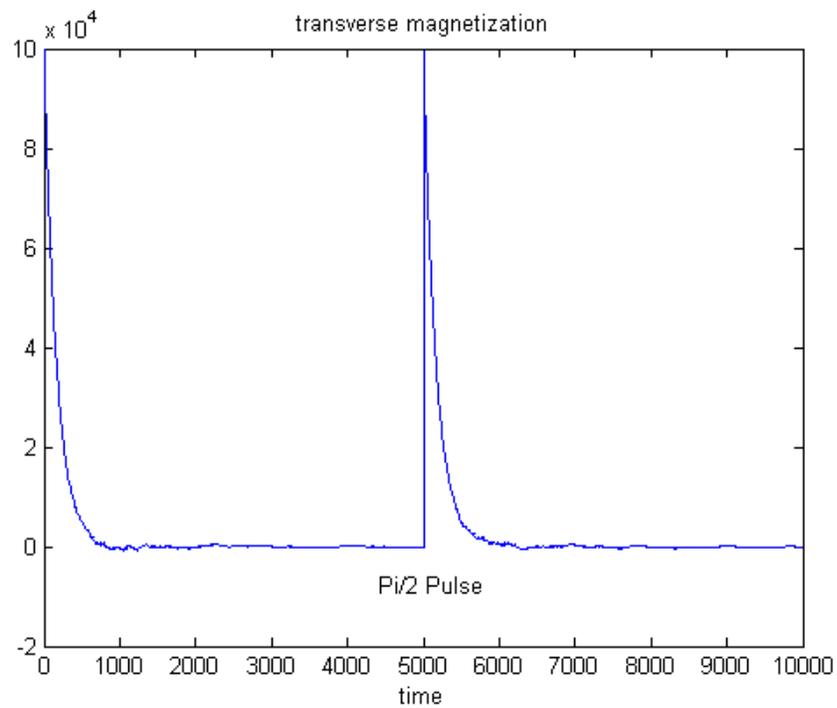

**Defacto Standard:** As the input to this computer simulation is probabilistic, the entire code, by definition, falls under the umbrella of Monte Carlo. Monte Carlo is defined as "a general term used to describe any method that utilizes a sequence of random numbers to perform calculations.[1]" However, finding other computer codes that model nuclear resonance is not so simple.

A search of the web for Monte Carlo reveals a large number of citations. However when cross referenced with Nuclear Magnetic Resonance Imaging, the number of citations goes down to 50,000 from 5,000,000. This drop to 1% of the citations is sizeable, but so is what is left over. Thus a tighter search refinement is needed to isolate the defacto standard[2].

Adding in the search criteria, "conditional branching" brings the number of citations to 5. This is a manageable number for reading. Of these five, two are a general overview of quantum computing, two are a detailed analysis of Monte Carlo code, and one is a conference agenda for the Molecular Graphics and Modeling Society. Performing other searches on Google, leads to a handful of helpful papers and Manuscripts that are downloadable from the web[3],[4].

With further research a similar conditional branching process as used in my code for motion is used in the Monte Carlo simulation available from TreeAge. However the subject of the Monte Carlo simulation is not motion, but serendipitously, it is gambling. For example a gambler who plays the lottery must first decide if he is going to buy a ticket. This additional stage of buying the ticket is similar to the conditional branching of colliding or not colliding. Only if the gambler goes ahead and buys a ticket does the second random branch

---

[1] Gustar et al, "Probabilistic Assessment of Structures, using Monte Carlo Simulation 2nd ed." Institute of Theoretical and Applied Mechanics, 2003.

[2] The Defacto Standard is the most widely used standard for a particular field of Monte Carlo.

[3] Remy, Tropres, Peoc'h, Farion, & Decorps. "In vivo NMR Imaging of microvascularization in mormal rat brain and in rat brain tumors" *Proc. Intl. Soc,. Mag. Reson. Med.* 8 (2000)

[4] Toumelin, Torres-Verin, Chen, Atlas. "Modeling of Multiple Echo-Time NMR Measurements for Complex Core Geometries and Multiphase Saturations" *Society of Petroleum Engineers* 2003 Vol 6 No 4

occur before the next time period when he can choose to buy a ticket again. If no collision occurs, then the code advances to the next stage without any event effecting the state.

The TreeAge and other Monte Carlo codes however, do not use the unique conditional branching that my adapted computer code uses. The reason why the code developed in this paper is viable is because it breaks down the individual collision events into case specific scenarios that then finishes by implementing the consequences of the outcome of a random number generator. There are other Monte Carlo codes available for modeling NMR, but they are more specific to implementing error analysis to determine the precision of an NMR-refined structure[5]. Thus the defacto standard for the Monte Carlo, by which to compare my code for motion, is available from TreeAge, but there is no defacto standard for the adapted code. This is because the analogy is not as exact when applied to the code for Magnetic Resonance Imaging because of the unique partitioning of the spins into the four basis states (even before a collision is simulated).

Basically the method developed here follows the Master Equation. Forming a set of probabilistic differential equations, the Master Equation is solved by integrating an ensemble of particles over all the possible states. Iterating through each state allows the unique code to touch on each spin and make sure that a possible collision is simulated. In the end, the code correctly models the relaxation of stimulated spins back to thermal equilibrium values.

---

[5]http://jet.uah.edu/~yaoys/tablemodel/mariland.ppt

```
N=100000;
Nz=N-Nt;
beta=.6;
Ntpos=beta*N;
Ntneg=Nt-Ntpos;
Nzpos=0; Nzneg=Nz-Nzpos;
T=10000;
alpha1=.001;
alpha2=.002;
for t=1:T
           nzn(t)=Nzpos-Nzneg;
           ntn(t)=Ntpos-Ntneg;
           Ntposold=Ntpos;
           Ntnegold=Ntneg;
           Nzposold=Nzpos;
           Nznegold=Nzneg;

             for n=1:Nzposold
             if rand<alpha1
                Nzpos=Nzpos-1;
                dn=ceil(rand-beta);
                Nzpos=Nzpos+not(dn);
                Nzneg=Nzneg+dn;
             end
             end
             for n=1:Nznegold
             if rand<alpha1
                Nzneg=Nzneg-1;
                dn=ceil(rand-beta);
                Nzpos=Nzpos+not(dn);
                Nzneg=Nzneg+dn;
             end
             end
             for n=1:Ntposold
             if rand<alpha1
                Ntpos=Ntpos-1;
                dn=ceil(rand-beta);
                Nzpos=Nzpos+not(dn);
                Nzneg=Nzneg+dn;
             elseif rand<alpha2
                Ntpos=Ntpos-1;
                dn=ceil(rand-.5);
                Ntpos=Ntpos+not(dn);
                Ntneg=Ntneg+dn;
             end
             end
             for n=1:Ntnegold
             if rand<alpha1
                Ntneg=Ntneg-1;
                dn=ceil(rand-beta);
                Nzpos=Nzpos+not(dn);
                Nzneg=Nzneg+dn;
             elseif rand<alpha2
                Ntneg=Ntneg-1;
                dn=ceil(rand-.5);
                Ntpos=Ntpos+not(dn);
                Ntneg=Ntneg+dn;
             end
           end
   end
```